\documentclass[twocolumn,showpacs,preprintnumbers,prl]{revtex4}
\usepackage{graphicx}
\usepackage{dcolumn}
\usepackage{bm}
\usepackage{graphicx}
\usepackage{dcolumn}
\usepackage{bm}

\def\be{\begin{equation}}
\def\ee{\end{equation}}
\def\bea{\begin{eqnarray}}
\def\eea{\end{eqnarray}}
\def\ba{\begin{array}}
\def\ea{\end{array}}

\def\vep{\varepsilon}

\def\0{$\Gamma_0$}

\def\p{\phi}

\def\t{\theta}

\def\s{\sigma}

\def\l{\lambda}

\begin{document}

\title{Superlens imaging theory for anisotropic nanostructured metamaterials with broadband all-angle negative refraction}
\author{W. T. Lu}
\email{w.lu@neu.edu}
\author{S. Sridhar}

\affiliation{Department of Physics and Electronic Materials Research Institute,
Northeastern University, Boston, MA 02115}
\date{\today }

\begin{abstract}
We show that a metamaterial consisting of aligned metallic nanowires in a dielectric matrix has strongly anisotropic optical properties. For filling ratio $f<1/2$, the composite medium shows two surface plasmon resonances (SPRs): the transverse and longitudinal SPR with wavelengths  $\l_t<\l_l$.
For $\l>\l_l$, the longitudinal SPR, the material exhibits $\mbox{Re}\ \vep_{||}<0 $, $\mbox{Re}\ \vep_{\bot}>0 $,  relative to the nanowires axis, enabling the achievement of broadband all-angle negative refraction and superlens imaging. An imaging theory of superlens made of these media is established. High performance systems made with Au, Ag or Al nanowires in nanoporous templates are designed and predicted to work from the infrared up to ultraviolet frequencies.
\end{abstract}

\maketitle

Since the demonstration of negative refraction (NR) \cite{Veselago} at microwave frequencies \cite{Shelby}, 
a variety of approaches have been described to observe the phenomenon at optical frequencies \cite{Shalaev07}. 
There are several reasons for the interest in NR, the most prominent application being 
the concept of perfect lens \cite{Pendry00} which can lead to sub-wavelength imaging beyond the diffraction limit. 
So far NR has been realized in periodic or quasiperiodic structures such as metamaterials 
\cite{Shelby} and photonic crystals \cite{Notomi,Luo,Parimi04}. 

As the frequency is increased to the optical spectrum, the structure size and the unit cell size shrink to 
nanometer dimensions. Important developments in nanofabrication do allow the fabrication of nanostructures down to 10 nm sizes over large areas. Using either top-down nanolithography or bottom-up self assembly, it is possible to fabricate aligned nanowires in dielectric matrices with large aspect ratios. 
For example, in alumina nanoporous templates \cite{Jessensky}, the pore diameter can be modified 
between 10 to 200 nm and the thickness 
can be a few nanometers up to 160 $\mu$m \cite{Lee}. Typical sizes of the pore diameter of 10 nm, pore distance of 50 nm can be easily obtained. Au nanowires synthesized inside the template make a uniform array of vertical nanowires arranged parallel 
to each other \cite{Sander,McMillian,Menon}. 

In this paper we show that such aligned nanowire structures in dielectric matrices constitute a class of indefinite index media with strongly anisotropic optical properties that can be used to achieve broadband all-angle NR (AANR) and superlens imaging. We show that these anisotropic media will have two surface plasmon resonances (SPR): a longitudinal SPR and a transverse SPR. For wavelength larger than that of the longitudinal SPR, these media are negative index metamaterials and can be used for superlens imaging \cite{Pendry00,Lu05} in the frequency range from the deep-infrared up to the ultraviolet. NR and superlens imaging are possible due to the anisotropic optical properties. These structures do not need to be periodic. Disordered structures can also be used for NR. 
Example systems are designed and demonstrated.

We consider a metal with ${\mbox Re}\>\vep_m<0$ embedded 
in an ambient medium with positive $\vep_a$.
In the long wavelength limit, one has the Bruggeman's effective medium theory (EMT)
\cite{Aspnes,Sihvola,Mackay}
\be
0=f{\vep_m-\vep_{\rm eff}\over \vep_m+D\vep_{\rm eff}}
+(1-f){\vep_a-\vep_{\rm eff}\over \vep_a+D\vep_{\rm eff}}. \label{EMT}
\ee
Here $f$ is the metal filling ratio and $D$ is a measure of the aspect ratio. 
The solution is
\be
\vep_{\rm eff}(D)={1\over 2D}\Big(\Delta\pm\sqrt{\Delta^2+4D\vep_a\vep_m}\Big)
\ee
with $\Delta=f(1+D)(\vep_m-\vep_a)+D\vep_a-\vep_m$.
The sign is chosen such that $\mbox{Im}\ \vep_{\rm eff}>0$.
For sphere inclusion, one has $D=2$ \cite{Aspnes}. 
For slab inclusion, $D=0$ and $\infty$ for the effective permittivity perpendicular and
parallel to the slabs, respectively.

\begin{figure}[tbp]
\center{
\includegraphics [angle=0, width=6cm]{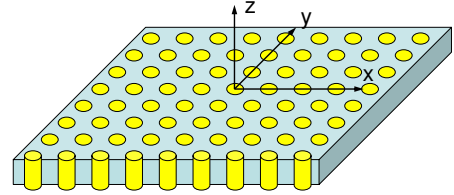}}
\vskip -2mm
\caption{(Color online) A slab of composite medium made of metal rods (yellow) 
embedded in a host medium (cyan).}
\label{fig1}
\end{figure}

For cylinder inclusion with cylindrical axis in the $z$-direction as shown in Fig. \ref{fig1},
one has 
\bea
\vep_x&=&\vep_y=\vep_{\rm eff}(1),\nonumber \\
\vep_z&=&\vep_{\rm eff}(\infty)=f\vep_m+(1-f)\vep_a.
\eea
There exists a minimal filling ratio
\be
f_{\rm min}=\vep_a/(\vep_a-\mbox{Re}\>\vep_m).
\ee
such that for $f>f_{\rm min}$, $\mbox{Re}\ \vep_z<0$. Also that if $f>1/2$, $\mbox{Re}\ \vep_{x,y}<0$.
If one desires $\mbox{Re}\ \vep_{x,y}>0$ but $\mbox{Re}\ \vep_z<0$, 
one should have $\mbox{Re}\ \vep_m<-\vep_a$ 
so that $f_{\rm min}<1/2$. 
We note that for the modelling of real systems, the values of $D$ can be
different from the ones we used \cite{note1}.

The physical meaning of $f_{\rm min}$ is the following. 
At this filling ratio which also corresponds to a fixed frequency or wavelength
$\l_l$ since $\vep_m$ is dispersive,
the composite medium has $\mbox{Re}\>\vep_z=0$ which gives strong absorption of the medium.
This frequency corresponds to the so-called longitudinal SPR \cite{Yu,Atkinson}.
For example for a Drude metal with $\vep_m=1-\l^2/\l_p^2$ and $\l_p$ the plasmon wavelength, 
one has $\l_l=\l_p[1+(f^{-1}-1)\vep_a]^{1/2}$. 
Thus $\l_l$ is very sensitive to the filling ratio $f$ and dielectric 
constant $\vep_a$ of the host medium.
The increase of the filling ratio results in a blue-shift of the longitudinal SPR.
The smaller the refractive index of the host medium, the shorter the longitudinal SPR $\l_l$.
High absorption is also expected for frequency at the so-called transverse SPR,
which is located around the surface plasmon wavelength $\l_{sp}$ ($\mbox{Re}\ \vep_m= -\vep_a$) 
and has very weak dependence on the filling ratio. 
For a Drude metal, there is a frequency range 
$\l_{t,+}<\l<\l_{t,-}$ with $\vep_m(\l_{t,\pm})
=\vep_a\{1-2(1-2f)^{-2}\pm4[f(1-f)]^{1/2}(1-2f)^{-2}\}$, such that 
$\mbox{Im}\ \vep_x>0$ and the medium shows strong absorption. 
Here $\vep_m(\l_{t,+})>-\vep_a$ and $\vep_m(\l_{t,-})<-\vep_a$.
For $f<0.1464$, one has $\l_{t,-}<\l_l$.

\begin{figure}[tbp]
\center{
\includegraphics [angle=0, width=7.8cm]{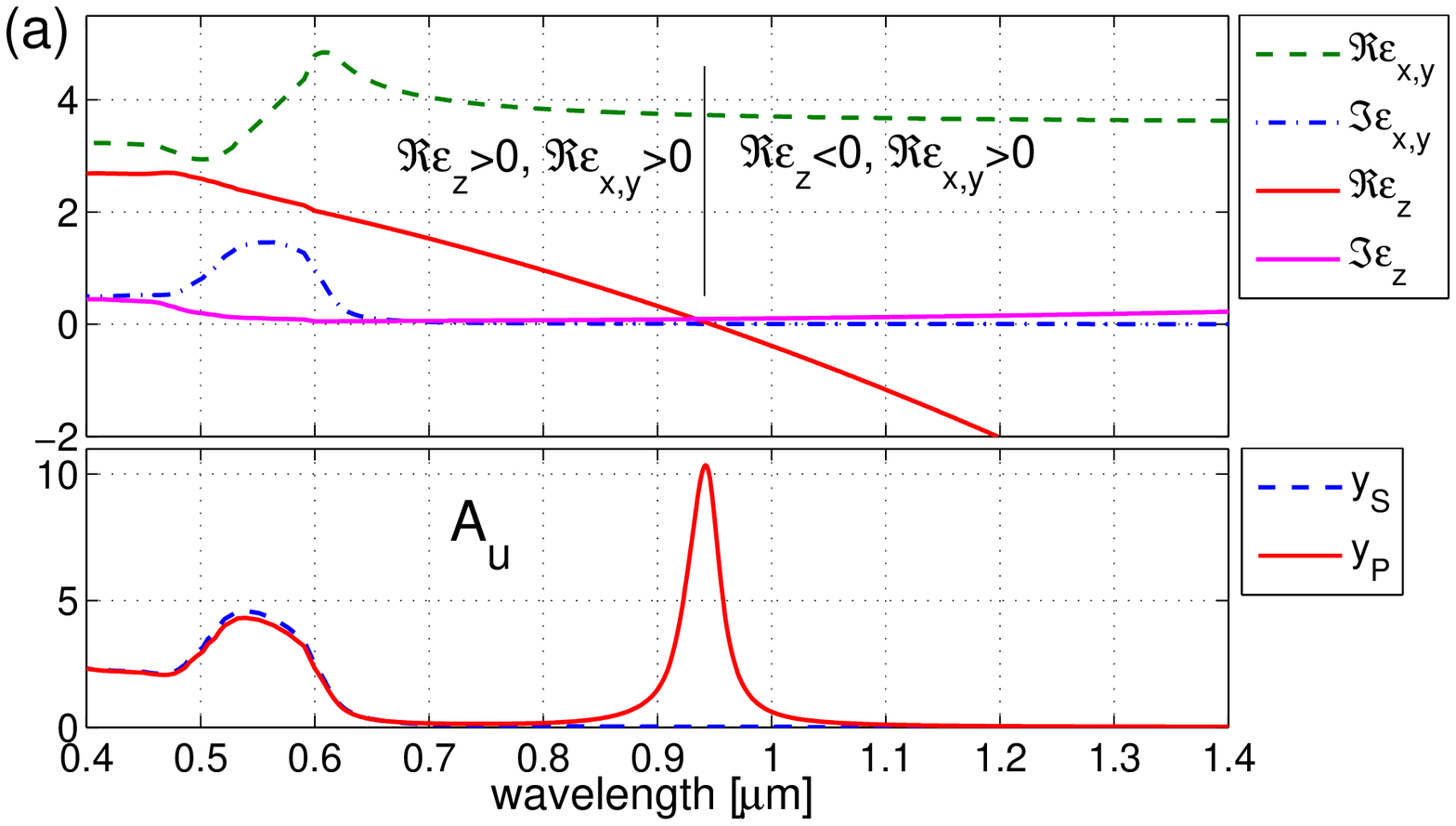}
\includegraphics [angle=0, width=7.8cm]{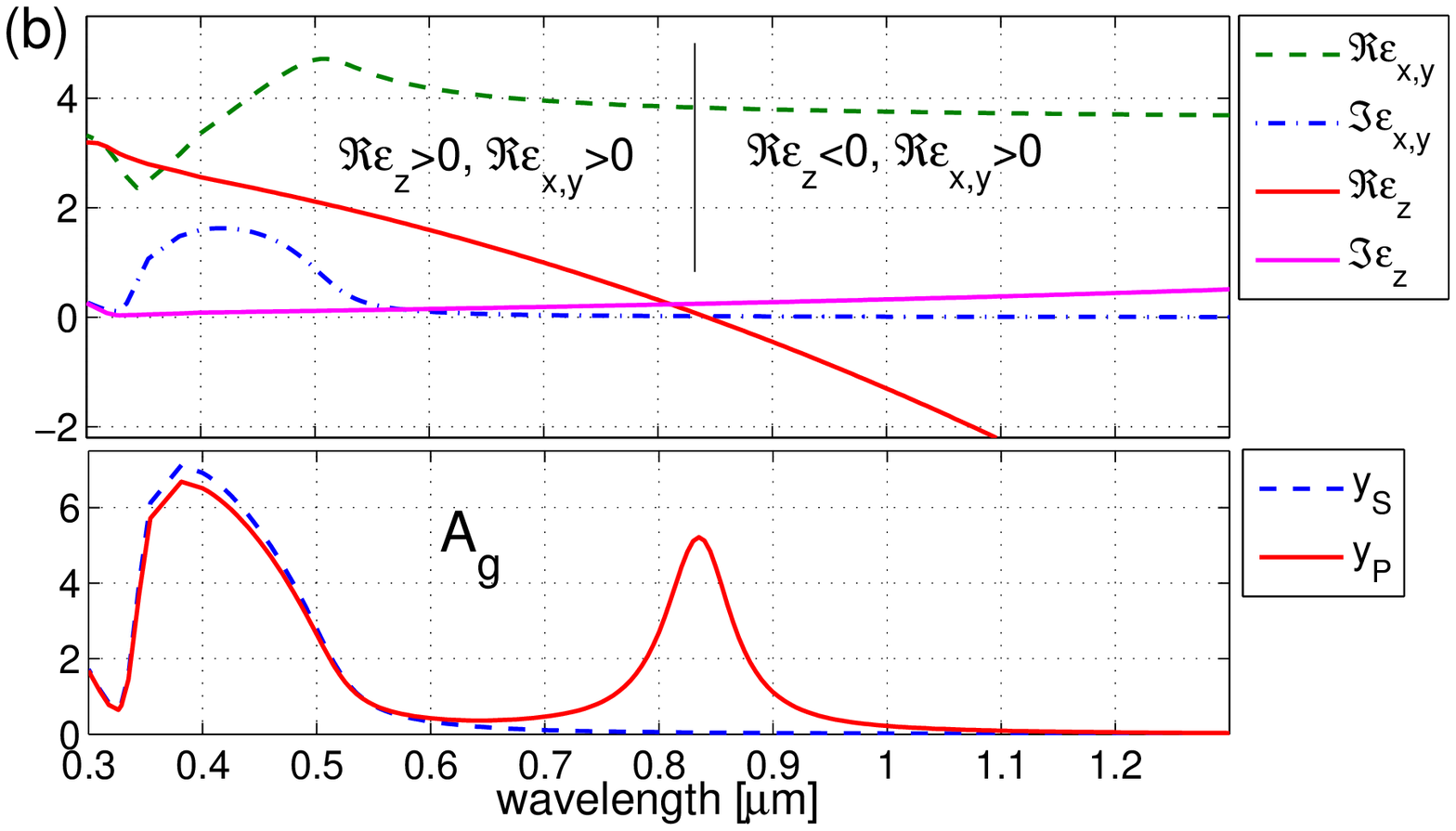}
\includegraphics [angle=0, width=7.8cm]{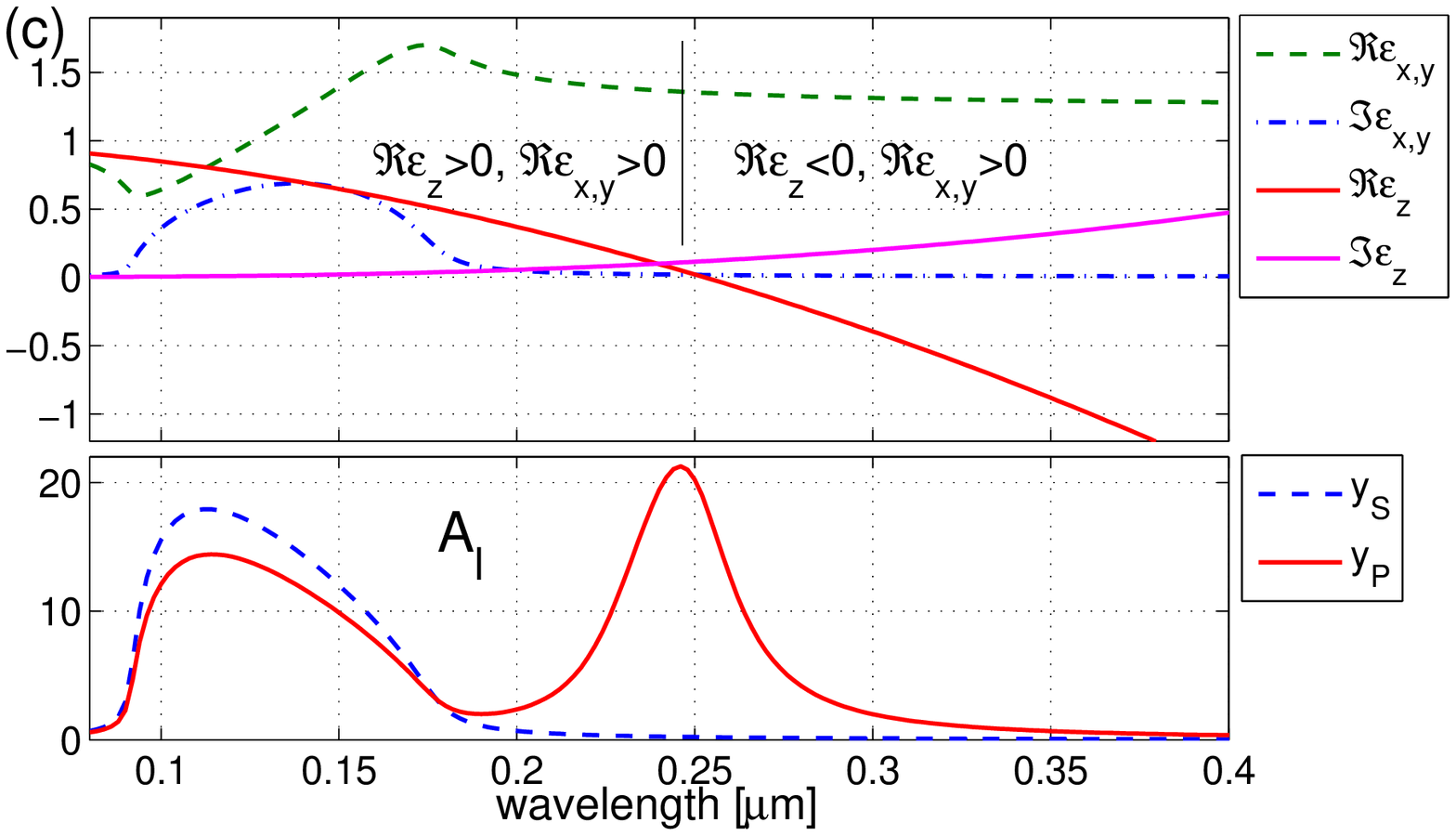}}
\vskip -3mm
\caption{(Color online) Effective permittivity and absorption spectra of metamaterials made of metallic 
nanowires embedded in dielectric masks
for $S$- and $P$-polarized waves with incident angle 25$^\circ$. 
The thickness of the metamaterials are all 500 nm.
(a) 8\% Au nanowires in alumina template, (b) 8\% Ag nanowires in alumina template,
(c) 10\% Al nanowires in air.
The dielectric constants $\vep_m$ are taken from taken Ref. \cite{Weaver}.
For simplicity we use $\vep_a=3$ for alumina in the infrared and the visible range.
Dashed lines are for the $S$-polarization and solid lines for the $P$-polarization.}
\label{fig2}
\end{figure}

For composite media with embedded Ag, Au, and Al nanowires, the effective permittivities and
the absorption spectra $y=\ln[(1-R)/T]$ are calculated and shown in Fig. \ref{fig2}. Here $R$ and $T$ are 
the reflection and transmission intensities of waves through a slab. 
The optical constants are taken from Ref. \cite{Weaver} and fitted with polynomials.
The absorption spectra clearly show the longitudinal SPRs
for the $P$-polarized waves.

\begin{figure}[tbp]
\center{
\includegraphics [angle=0, width=7.6cm]{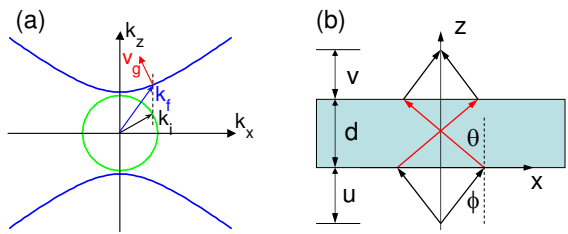}}
\caption{(Color online) (a) Illustration of negative refraction and (b) superlens imaging
 of $P$-polarized waves by a slab of indefinite medium with $\mbox{Re}\>\vep_z<0$ and $\mbox{Re}\>\vep_x>0$.
The blue line is the EFS in this medium and the green line is the corresponding EFS in the air.
An incident plane wave with ${\vec k}_i$ will be refracted as ${\vec k}_f$ 
which has negative group refraction (red arrow). The superlens has a lens equation $u+v=\s d$.}
\label{fig3}
\end{figure}

When the metamaterial has $\mbox{Re}\ \vep_z<0$ and $\mbox{Re}\ \vep_{x,y}>0$, 
this so-called indefinite medium \cite{Smith03} has unusual wave refraction phenomena and 
can be used for NR and superlens imaging for incident 
waves along the nanowire axis.
We consider a slab of such medium 
whose surface is along the $x$-axis and 
surface normal is along the $z$-axis as shown in Fig. \ref{fig3}(b). 
We assume the relative permeability is of unity.
For the $P$-polarization with the magnetic field in the $y$-direction
and the electric field in the $xz$-plane, the dispersion is
$k_z^2=\vep_xk_0^2-\vep_xk_x^2/\vep_z$.
Here $k_0=2\pi/\l$ the wave number in free space.
When $\mbox{Re}\ \vep_z<0$ and $\mbox{Re}\ \vep_{x,y}>0$, 
the equi-frequency surface (EFS) is hyperbolic instead 
of elliptic as shown in Fig. \ref{fig3}(a).
For this medium, it is more meaningful to discuss the energy flow.
The group velocity refraction is governed by \cite{Lu05}
\be
\tan \t=-\s\tan\p.
\ee
Here $\p$ is the angle for the incident group velocity and $\t$ is that for
the refracted group velocity (see Fig. \ref{fig3}(b)).
The material property $\s$ is defined and evaluated as \cite{Lu05}
\be
\s\equiv-{dk_z\over dk_{0z}}=-{\sqrt{\vep_x}\over \vep_z}{\sqrt{k_0^2-k_x^2}\over \sqrt{k_0^2-k_x^2/\vep_z}}.
\ee
Here $k_{0z}=\sqrt{k_0^2-k_x^2}$.
One has $\s>0$ for all propagating waves, 
thus AANR \cite{Luo} can be realized in this medium.
The group refractive index $n_{\rm eff}$ is related to $\s$ through 
$n_{\rm eff}\sin\t=\sin\p$.
One has $n_{\rm eff}\sim -\s^{-1}$.
For small $k_x$ the EFS can be approximated elliptically by
$k_z\simeq \kappa-\s_0k_{0z}$
with $\s_0=-\sqrt{\vep_x}/\vep_z>0$ and $\kappa=\sqrt{\vep_x}k_0(1-1/\vep_z)$.

When $\s$ is constant, then the phase across the lens $\Phi_z = k_{0z}(u+v)+k_zd =\kappa d$ 
is stationary and an image will be formed
without aberration. Here $u$, $v$, and $d$ obey the equation $u + v = \s d$ (Fig. \ref{fig3}(b)). 
This is the lens equation for a generalized superlens. 
In this case the refractive index $n_{\rm eff}$ is angle-dependent, and one can achieve 
``perfect focusing" without an optical axis, as discussed in Ref. \cite{Lu05}.
Note that the Veselago lens has $\s = 1$, where the EFS is circular.

In the present anisotropic metamaterial, $\s$ is angle-dependent and not a constant, 
because the EFS is hyperbolic and not elliptic \cite{Lu05}; hence the lens has caustics, 
and the image is not ``perfect". 
Nevertheless, a high quality image can be formed
by the lens with $u+v=\s_{\rm eff} d$ and $\s_{\rm eff}<\s_0$. 
Furthermore, though the nonlocal effect \cite{Pokrovsky02,Belov03,Silveirinha06,Simovski07,Elser} 
on effective permittivity indicates the limitation of Bruggeman's EMT, it will render the
EFS to be more elliptic than hyperbolic, thus can reduce the caustics.

The composite medium with cylinder inclusion can be used for NR 
and superlens imaging in three-dimensional free space
for frequencies below the surface plasmon frequency.
These metamaterials do not support surface waves. 
The enhancement of subwavelength imaging resolution is still possible \cite{Ono,Silveirinha07,Shvets07}. 
If the lens is curved, one may be able to 
use it as a magnifying hyperlens \cite{Jacob,Engheta06}.
The currently studied multilayered structures \cite{Jacob,Engheta06,Liu07,Shin,Fan,Pendry06,Hoffman}
for NR, superlens, and hyperlens are two-dimensional
reductions of these structures.
The filling ratio $f=1/2$ is special for multilayered metal-dielectric structures.
At this filling ratio, $\mbox{Re}\ \vep_x$ and  $\mbox{Re}\ \vep_z$ will always have the opposite signs.
This has been utilized to realize magnifying hyperlens \cite{Jacob,Engheta06,Liu07}.
Naturally available anisotropic dielectric crystals may be used to achieve NR \cite{ZhangY}
but can not be used for superlens imaging.

\begin{figure}[tbp]
\center{
\includegraphics [angle=0, width=7.6cm]{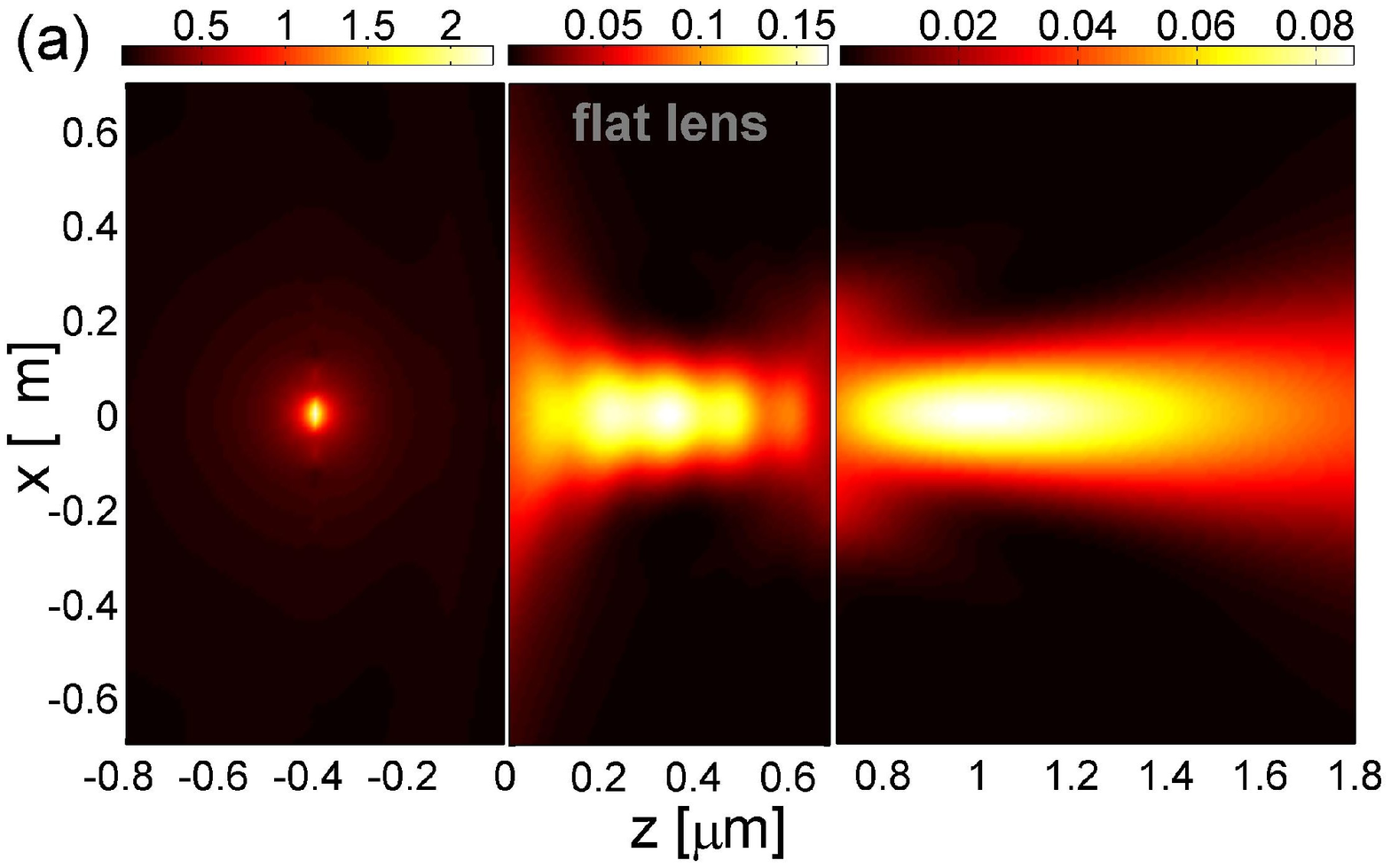}
\hskip 3mm
\includegraphics [angle=0, width=7.6cm]{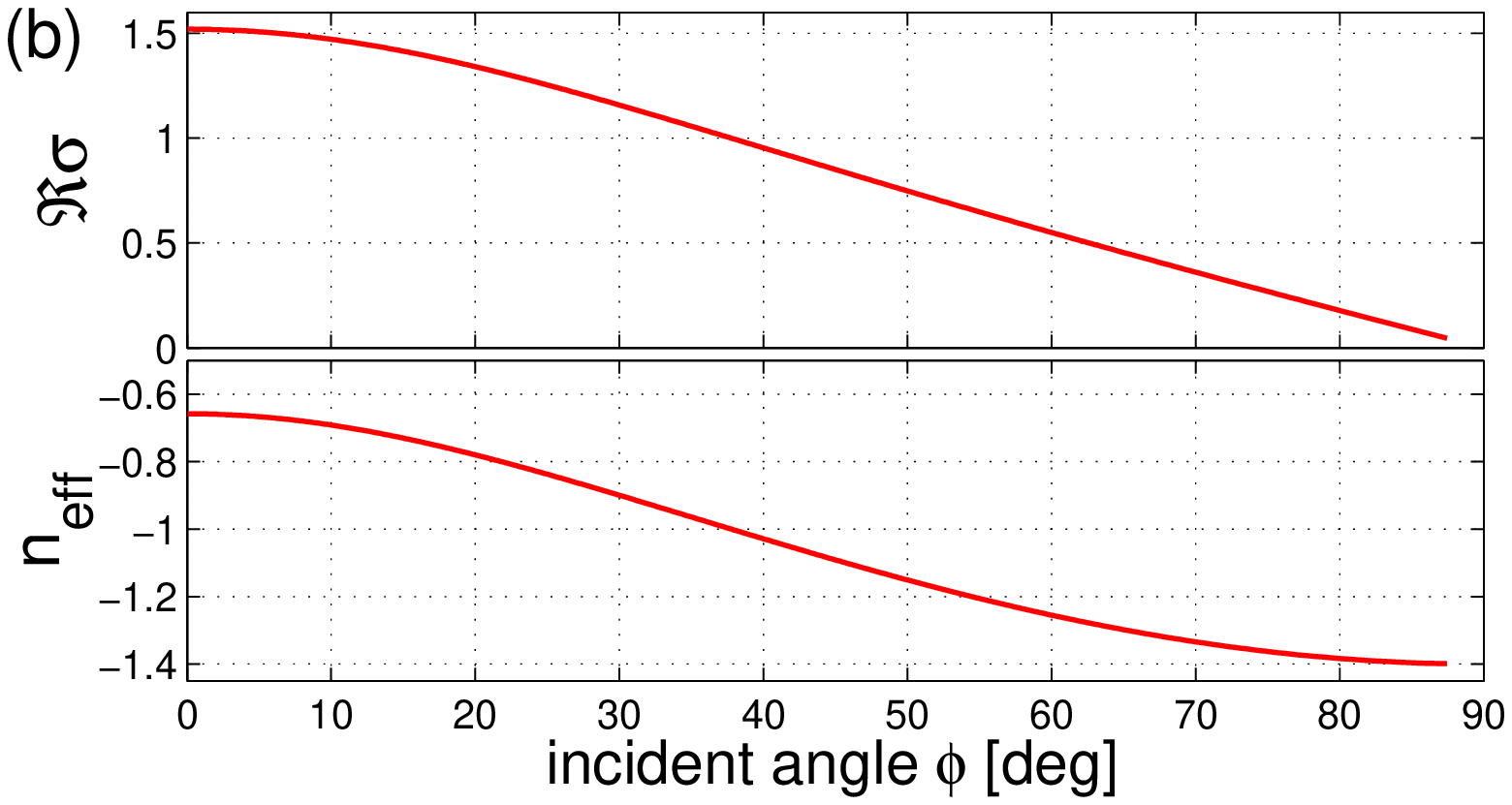}}
\vskip -2mm
\caption{(Color online) (a) Imaging of a superlens with $\vep_x=1.301 + 0.010i$, $\vep_z=-0.647 + 0.258i$,
and thickness $d=0.7\ \mu$m, at $\l=326.3$ nm. The point source is at $u=0.40\ \mu$m and
a focus is obtained at $v=0.30\ \mu$m. The lens has $\s_{\rm eff}=1.00$. 
Plotted is the intensity of the magnetic field which is in the $y$-direction. 
Evanescent waves of the source are included up to $k_x/k_0=3$. (b) Lens property $\s$ and $n_{\rm eff}$
as functions of the incident angle.}
\label{fig4}
\end{figure}

Anisotropic metamaterials with embedded Au, Ag, Al nanowires can be used for superlens 
imaging in the infrared, visible, and ultraviolet, respectively. 
For example for Al at $\l=326.3$ nm, $\vep_m=-15.468 + 2.575i$ \cite{Weaver},
a 10\% filling ratio of Al nanowires in air
 gives $\vep_x=1.301 + 0.010i$ and $\vep_z=-0.647 + 0.258i$. 
A lens made of a flat slab of such medium has $\s_0=1.52$
and can have a maximum thickness 11.9 $\mu$m.
The imaging effect of a point source by such a superlens is shown in  Fig. \ref{fig4}(a). 
The angle-dependent lens property $\s$ as shown in Fig. \ref{fig4}(b) leads to the presence of caustics
which can be reduced if multiple lenses are used.

For the propagating waves within the $xy$ plane, NR and superlens imaging 
can be realized in a finite slab of such an anisotropic medium. 
In this case for the $P$-polarized waves, one has 
$k_x=\sqrt{\vep_z}\sqrt{k_0^2-k_z^2/\vep_x}$. For $\l>\l_l$, $\mbox{Re}\ \vep_z<0$, 
a free-suspending slab will support guided waves in the $xy$ plane if $\sqrt{\mbox{Re}\ \vep_x}>1$
and $k_z>\sqrt{\mbox{Re}\ \vep_x}k_0$.
These guided waves are backward waves with $\mbox{Re}\ k_x<0$. 
In this geometry, surface waves can be formed
which can lead to subwavelength imaging resolution.
We point out that there is no need to sandwich this medium by perfect conductor 
waveguide plates as considered in Ref. \cite{Narimanov}.

For the $S$-polarization, the medium with cylinder inclusion is isotropic
with positive effective permittivity.
The dispersion is given by
$k_z^2=\vep_yk_0^2-k_x^2$. No NR can be realized for this polarization.

There are two strategies to realize $\mbox{Re}\>\vep_z<0$ and $\mbox{Re}\>\vep_x>0$
depending on the wavelength $\l>\l_{sp}$ or $\l<\l_{sp}$.
Clearly sphere inclusion can not while the cylinder or slab inclusion 
will lead us to the desired anisotropy. 
For $\l>\l_{sp}$, the cylinder axis should be along the $z$-axis
which we have discussed in this paper.
The slab inclusion must be realized as a metallic grating as in Ref. \cite{Fan}. 
In this case one has $\mbox{Re}\>\vep_{y,z}<0$
and $\mbox{Re}\>\vep_x>0$. Though loss is low in $\vep_x$, NR is limitted to the $xz$-plane.
We point out that the EMT theory gives a very simple explanation for
the broadband AANR in the metallic grating \cite{Fan}.
Furthermore, the EMT is more accurate than the coupled-wave 
theory used in Ref. \cite{Fan}. We also point out that 
our numerical simulations indicate that AANR does not require the metallic grating to be periodic.

For $\l_p<\l<\l_{sp}$, indefinite medium can be realized if
the cylinder axis is in the $xy$-plane for the cylinder inclusion.
If the cylinder axis is along the $x$-axis,
one has $\vep_x=\vep_{\rm eff}(\infty)$ and 
$\vep_{y,z}=\vep_{\rm eff}(1)$.
For these wavelengths $-\vep_a<\mbox{Re}\ \vep_m<0$, 
one should have $1/2<f<\vep_a/(\vep_a-\mbox{Re}\>\vep_m)$,
high loss will be expected for $\vep_x$. However the slab inclusion which is
examplified by the multilayered metal-dielectric structures considered in Ref. \cite{Shin,Pendry06,Hoffman}, 
will have $\mbox{Re}\>\vep_z<0$ and 
$\mbox{Re}\>\vep_{x,y}>0$ 
for $-\vep_m/(\vep_a-\mbox{Re}\>\vep_m)<f<\vep_a/(\vep_a-\mbox{Re}\>\vep_m)$ with low loss. 
For these structures \cite{Shin,Pendry06,Hoffman}, Bruggeman's EMT may not be very precise 
to calculate the effective permittivity, 
but our imaging theory predicts that they are able to focus.

In summary, we have established an imaging theory of superlens
made of nanoporous dielectric templates embedded with
metallic nanowires with strong optical anisotropy. 
The effective permittivity of these media is obtained by Bruggeman's EMT.
Our theory is successfully used to give simple explanation of recent 
predictions of NR and superlens focusing in multilayered \cite{Fan,Shin,Pendry06,Hoffman} 
and 2D structures \cite{Ono,Silveirinha07}.
For frequencies lower than the longitudinal SPR frequency, 
these media are indefinite media \cite{Smith03}.
The existence of $\mbox{Re}\ \vep_z<0$ and $\mbox{Re}\ \vep_{x,y}>0$ is manifested by the fact 
that there is a band cutoff
for the TM modes and no band cutoff for the TE modes of 2D metal-dielectric photonic crystals. 
The homogenization \cite{note1,HuX} including the nonlocal effect 
\cite{Pokrovsky02,Belov03,Silveirinha06,Simovski07,Elser}
of 2D photonic crystals
provides direct and more accurate ways to calculate the effective
indices beyond EMT.
Unlike resonant metamaterials which are sensitive to disorder \cite{Gorkunov}, these nanowire media
has large tolerance on disorder.
These extremely anisotropic metamaterials are broadband and can be used for NR, superlens, 
and hyperlens applications in 3D for frequencies from infrared up to ultraviolet.

We thank L. Menon, D. Casse and R. Banyal for useful discussions.
This work was supported by the Air Force Research
Laboratories, Hanscom through FA8718-06-C-0045 and
the National Science Foundation through PHY-0457002.


\begin{thebibliography}{9}

\bibitem{Veselago} V. G. Veselago, 
Sov. Phys. Usp. {\bf 10}, 509 (1968).

\bibitem{Shelby} R. A. Shelby, D. R. Smith, and S. Shultz, 
Science {\bf 292}, 77 (2001).

\bibitem{Shalaev07} V. M. Shalaev, Nat. Photon. {\bf 1}, 41 (2007).

\bibitem{Pendry00} J. B. Pendry, 
Phys. Rev. Lett. {\bf 85}, 3966 (2000).

\bibitem{Notomi} M. Notomi, Phys. Rev. B {\bf 62}, 10696 (2000).

\bibitem{Luo} C. Luo, S. G. Johnson, J. D. Joannopoulos, and J. B. Pendry, 
Phys. Rev. B {\bf 65}, 201104 (2002).

\bibitem{Parimi04} P. V. Parimi {\it et al}., Phys. Rev. Lett. \textbf{92}, 127401
(2004); P. V. Parimi {\it at al}., 
Nature {\bf 426}, 404 (2003).

\bibitem{Jessensky} O. Jessensky, F. M\"uller, and U. G\"osele, 
Appl. Phys. Lett. {\bf 72}, 1173 (1998).

\bibitem{Lee} W. Lee et al., Nat. Mat. {\bf 5}, 741 (2006).

\bibitem{Sander} M. S. Sander and L.-S. Tan, Adv. Func. Mat. {\bf 13}, 393 (2003);
O. Rabin {\it et al}., Adv. Func. Mat. {\bf 13}, 631 (2003).

\bibitem{McMillian} B. G. McMillian {\it et al}., Appl. Phys. Lett. {\bf 86}, 211912 (2005).

\bibitem{Menon} L. Menon {\it et al}., preprint (2007).

\bibitem{Lu05} W. T. Lu and S. Sridhar, Opt. Exp. \textbf{13}, 10673 (2005).

\bibitem{Aspnes} D. E. Aspnes, Am. J. Phys. \textbf{50}, 704 (1982). 

\bibitem{Sihvola} A. Sihvola, {\it Electromagnetic mixing formulas and applications},
Institute of Electrical Engineers, London (1999).

\bibitem{Mackay} T. G. Mackay and A. Lakhtakia, Opt. Comm. {\bf 234}, 35 (2004).

\bibitem{note1} The homogenization of 2D photonic crystals gives 
$D\sim 1$ and $D\gg 1$ for $\vep_{x,y}$ and $\vep_z$, respectively.
There is no much difference in our subsequent discussions if more accurate expression is used.
For homogenization of photonic crystals, see P. Halevi, A. A. Krokhin, and J. Arriaga,
Phys. Rev. Lett. {\bf 82}, 719 (1999); {\it ibid}. {\bf 86}, 3211 (2001);
A. A. Krokhin and E. Reyes, {\it ibid}. {\bf 93}, 023904 (2004).

\bibitem{Yu} Y.-Y. Yu {\it et al}., 
J. Phys. Chem. B {\bf 101}, 6661 (1997).

\bibitem{Atkinson} R. Atkinson {\it et al}., Phys. Rev. B {\bf 73}, 235402 (2006).

\bibitem{Weaver} J. H. Weaver et al., {\it Optical properties of metals}, 
Fachinformationszentrum, Karlsruhe, Germany (1981).


\bibitem{Smith03} D. R. Smith and D. Schurig,
Phys. Rev. Lett. {\bf 90}, 077405 (2003);
D. R. Smith, P. Kolinko, and D. Schurig,
J. Opt. Soc. Am. B {\bf 21}, 1032 (2004).

\bibitem{Pokrovsky02} A. L. Pokrovsky and A. L. Efros, Phys. Rev. B {\bf 65}, 045110 (2002).

\bibitem{Belov03} P. A. Belov {\it et al}., 
Phys. Rev. B {\bf 67}, 113103 (2003); P. A. Belov and C. R. Simovski, Phys. Rev. E {\bf 72}, 026615 (2005).

\bibitem{Silveirinha06} M. G. Silveirinha, Phys. Rev. E {\bf 73}, 046612 (2006).

\bibitem{Simovski07} C. R. Simovski and S. A. Tretyakov, 
Phys. Rev. B {\bf 75}, 195111 (2007).

\bibitem{Elser} J. Elser {\it et al}., 
Appl. Phys. Lett. {\bf 90}, 191109 (2007).

\bibitem{Ono} A. Ono and S. Kawata, Phys. Rev. Lett. {\bf 95}, 267407 (2005).

\bibitem{Silveirinha07} M. G. Silveirinha, P. A. Belov, and S. R. Simovski,
Phys. Rev. B {\bf 75}, 035108 (2007).

\bibitem{Shvets07} G. Shvets, S. Trendafilov, J. B. Pendry, and A. Sarychev, Phys. Rev. Lett. {\bf 99}, 053903 (2007).

\bibitem{Jacob} Z. Jacob, L. V. Alekseyev, and E. Narimanov, Opt. Exp. {\bf 14}, 8247 (2006).

\bibitem{Engheta06} A. Salandrino and N. Engheta,
Phys. Rev. B {\bf 74}, 075103 (2006).

\bibitem{Liu07} Z. Liu {\it et al}., Science {\bf 315}, 1686 (2007);
I. I. Smolyaninov, Y.-J. Hung, and C. C. Davis, Science {\bf 315}, 1699 (2007).

\bibitem{Fan} X. Fan {\it et al}., Phys. Rev. Lett. {\bf 97}, 073901 (2006).

\bibitem{Shin} H. Shin and S. Fan, 
Appl. Phys. Lett. {\bf 89}, 151102 (2006).

\bibitem{Pendry06} B. Wood, J. B. Pendry, and D. P. Tsai, Phys. Rev. B {\bf 74}, 115116 (2006).

\bibitem{Hoffman} A. J. Hoffman {\it et al}., Nat. Mat., doi:10.1038/nmat2033.

\bibitem{ZhangY} Y. Zhang, B. Fluegel, and A. Mascarenhas, Phys. Rev. Lett. {\bf 91}, 157404
(2003).



\bibitem{Narimanov} V. A. Podolskiy and E. Narimanov, Phys. Rev. B {\bf 71}, 201101(R) (2005).

\bibitem{HuX} X. Hu {\it et al}., Phys. Rev. Lett. {\bf 96}
223901 (2006). 

\bibitem{Gorkunov} M. V. Gorkunov {\it et al}., 
Phys. Rev. E {\bf 73}, 056605 (2006).

\end{thebibliography}
\end{document}